\documentstyle[aps,12pt]{revtex}

\begin{document}
\title{Comment on ''Implementation of the LDA+U method using the
full-potential linearized augmented plane-wave basis'' }
\author{R. J. Radwanski}
\address{C{enter for Solid State Physics, S}$^{{nt}}${\ Filip 5, 31-150 Krak%
\'{o}w, }\\
{Inst. of Physics, Pedagogical University, 30-084 Krak\'{o}w, POLAND. }}
\author{Z.Ropka}
\address{{Center for Solid State Physics, S}$^{{nt}}${\ Filip 5, 31-150 Krak%
\'{o}w,}\\
{\ POLAND.}}
\maketitle

\begin{abstract}
\end{abstract}


In Ref. 1 Shick et al. have applied the combination of the local density
approximation plus Hubbard U (LDA+U) total-energy functional with the
full-potential linearized augmented plane-wave method to NiO and Gd.

By this Comment we would like to point out that the authors of Ref. 1 have
ignored a recent paper, i.e. Ref. 2. The importance of this paper is related
with the experimental evaluation of the magnetic moment of
single-crystalline NiO for 2.2 $\mu _B$ at 300 K. This value, being much
larger than the assumed in Ref. 1 value of 1.70 $\mu _B$ proves that the
calculated result for the magnetic moment of NiO substantially disagrees
with the experimental observation.

The second quantity calculated in Ref. 1 was an energy gap, attributed to
the insulating gap. The authors of Ref. 1 has got a large gap of 3.38 eV, 8
times larger than in LDA\ calculations. To get confidence to this result
authors should explain the reason for the dramatic increase in the exchange
splitting of the Ni d states. As far as the parameters of the theory are not
experimentally verified it is only a mere speculation. We fully agree with
the authors of Ref. 1 that ''the magnetic moment is a fundametal test'' for
the theory. We would like to point out that the authors of Ref. 1 can
calculate only the spin moment.

The experiment of Ref. 2 allows the determination of the spin and the
orbital momenta, m$_{s}$=1.90$\pm $0.10 $\mu _{B}$ and m$_{o}$=0.32$\pm $%
0.05 $\mu _{B}$. These values are measured at 300 K - one can expect at 0 K
values larger by at least 15\%. The observation of such substantial orbital
moment proves the uselessness of the theory of Ref. 1 to real systems
because the theory of Ref. 1 cannot calculate the orbital moment.

Moreover, we would like to point out that theoretical calculations without
thermodynamics are hardly of interest, both for experiment as well as for
theoretical understanding.

Authors of the commented paper should know Ref. 2 as Ref. 2 has been printed
1.04.1998 whereas the Ref. 1 has been submitted 1.04.1999, i.e. exactly 1
year later. Surely, the referee should put the authors' attention to this
paper.

In conclusion, the authors of Ref. 1 have missed new experimental results of
Ref. 2 that disqualify their theoretical result about the magnetic moment of
NiO. Moreover, despite of the very complex name of the approach, the
approach of Ref. 1 suffers substantial limitations related with the
neglection of the orbital magnetism and the unability to provide
thermodynamics. These deficiencies cause that the results are incompatible
to the reality. Here we can only mention our atomic-like approach, that
gives surprisingly good results for LaCoO$_{3}$ [3,4], FeBr$_{2}$ [5] and
NiO [6-8]. Our approach yields the discrete energy spectrum for the d states
in NiO in contrary to the continuum energy spectrum of Ref. 1 (see Fig. 2).
It provides the spin and orbital moments at 0 K as 1.99 and 0.54 $\mu _{B}$
[4,5], respectively, in surprisingly good agreement with experimental data
of Ref.2.

At the end we would like to say that we think that authors should have
scientific rights to publish their results as Science develops only in the
open discussion. The only problem appears when the Editor starts to
manipulate the Science by unresponsible rejecting of paper by means of, for
instance, unscientific statements like ''paper is not suited for
publication'' or ''not of wide interest'' without trying to get the clue of
the scientific controversy between referee and the author. This manipulation
unfortunatelly often goes with the help of referees who in anonymous reports
are scientifically unresponsible.

\footnote{%
e-mail: sfradwan@cyf-kr.edu.pl; URL: http://css-physics.edu.pl}

\end{document}